\documentclass[10pt,twoside]{article}
\usepackage[pdftex]{color,graphicx}
\usepackage[ascii]{inputenc}
\usepackage[T1]{fontenc}
\usepackage[english,spanish]{babel}
\usepackage{amsmath,amssymb,amsfonts,textcomp}
\usepackage{color}
\usepackage{calc}
\usepackage{multicol}
\usepackage{array,longtable}
\usepackage{hyperref}
\usepackage{graphicx,subfigure,epsfig,amsmath,setspace,amsfonts}
\hypersetup{colorlinks=true, linkcolor=blue, filecolor=blue, pagecolor=blue, urlcolor=blue}

\newcommand\textsubscript[1]{\ensuremath{{}_{\text{#1}}}}
\makeatletter
\newcommand\ps@Standard{}%
\renewcommand\@oddhead{}%
\renewcommand\@evenhead{}%
\renewcommand\@oddfoot{}%
\renewcommand\@evenfoot{}%
\makeatother
\everymath{\displaystyle}
{\selectlanguage{english} 
\title{A procedure to Estimate the Fractal Dimension of Waveforms\footnote{\selectlanguage{english}This work was originally published in Complexity International, an on-line journal. Since its publication, the journal became inactive, and several symbols became corrupted on-line. According to Complexity International, the author is always the owner of the article's copyright as long as the identification of Complexity International appears in the copies. Several typos are corrected in this version uploaded to \textit{\textbf{arXiv}}.}}

\author{Carlos Sevcik
\footnote{\selectlanguage{english}Email: csevcik@ivic.gob.ve}}
}

\begin{document}

	\begin{figure*}
		\includegraphics[width=98pt]{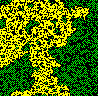}	

		\begin{Huge}
		\textbf{Complexity International}
		\end{Huge}
		\newline
		\begin{LARGE}
			ISDN 1320-0683
		\end{LARGE}
		\newline
		Source: http://www.complexity.org.au/ci/vol05/sevcik/sevcik.html
		Vol 5: Copyright 1998
		\newline
		Recived: 16/10/1997
		Accepted: 16/1/1998

	\end{figure*}
	\selectlanguage{english}\maketitle

	{\selectlanguage{english}
	\itshape
	
	Laboratory on Cellular Neuropharmacology
	Centro de Biof\'isica y Bioqu\'imica
	Instituto Venezolano de Investigaciones Cient\'ificas (IVIC)
	Apartado 21827, Caracas 1020A
	Venezuela.}

	\bigskip
	{\selectlanguage{english}
	\begin{abstract}
		
		I derived a method for calculating the approximate fractal dimension (\textit{D}) from a set of \textit{N} values \textit{y} sampled from a waveform between time zero and \textit{t\textsubscript{max}} with sampling interval ${\delta}$. The waveform was subjected to a double linear transformation that maps it into a unit square. The normalized abscissa of the square is ${x_{i}^{\text{*}}}$ and the normalized ordinate is ${y_{i}^{\text{*}}}$, both of them defined as
		
		\begin{equation*}
			\begin{gathered}
				x_{i}^{\text{*}}=\frac{x_{i}}{x_{\mathit{max}}} \\
				\\ y_{i}^{\text{*}}=\frac{y_{i}-y_{\mathit{min}}}{y_{\mathit{max}}-y_{\mathit{min}}}
			\end{gathered}
		\end{equation*}
		
		where $x_{max}$ is the maximum ${x_i}$ and $y_{min}$ and $y_{max}$ are the minimum and maximum $y_i$. The fractal dimension of the waveform ($\Phi$) is then approximated by \textit{D} as
		
		\begin{equation*}
			\Phi \approx D=1+\frac{\ln (L)}{\ln (2\cdot N\text{{\textquotesingle}})} 
		\end{equation*}

		where \textit{L} is the length of the curve in the unit square and ${N'=N -1}$.
	
	\end{abstract}}
	
	\section[1 Introduction]{1 Introduction}

		Studying living systems as non-linear dynamic systems (chaotic systems as they are commonly called) is of increasing interest to medicine and biology [8]. The fractal dimension is one possible parameter that characterizes chaotic systems, and the analysis of time series is one of the most common means to find the fractal dimension from observables [8]. 
		
		The analysis of time series is also interesting $\textit{per se}$. However, this analysis may deal with complicated properties such as regularity, complexity or spatial extent [17, 19, 20]. A good example may be found in two series presented by Pincus $\textit{et al.}$ [20] to illustrate the complexities of studying heart beat in normal and ill humans, these are:
		
		{\centering\selectlanguage{english}
		90, 70, 90, 70, 90, 70, 90, 70, 90, 70, 90, 70, 90, 70, 90, 70,...
		\par}
		
		and
		
		{\centering\selectlanguage{english}
		90, 70, 70, 90, 90, 90, 70, 70, 90, 90, 70, 90, 70, 70, 90, 70,...
		\par}
		
		These two series have the same mean, median and variance and the two values (90 or 70) have the same probability of occurring: \textonehalf. Rank statistics also fails to distinguish among these two series. Yet, the two series are different; in the first one we always know the next outcome with absolute certainty. In the second series, we only know that the next outcome will be either 90 or 70, but our guess will be wrong in 50\% of the cases.

{\selectlanguage{english}		
		The term waveform applies to the shape of a wave, usually drawn as instantaneous values of a periodic quantity ${\textit{versus}}$ time. Any waveform is an infinite series of points. Aside of classical methods such as moment statistics and regression analysis, properties such as the Kolmogorov-Sinai entropy [9], the apparent entropy [20] and the fractal dimension [15] have been proposed to tackle the problem of pattern analysis of waveforms. The fractal dimension may convey information on spatial extent (convolutedness or space filling properties) and self-similarity (the ability to remain unchanged when the scale of measurement is changed) and self affinity [2]. Unfortunately, although rigorous methods to find out the fractal dimension exist [10, 11, 1], their usefulness is severely limited since they are computer intensive and their evaluation is time consuming. In two-dimensional spaces, waveforms are planar curves.

		The fractal analysis of waveforms was introduced by Katz [15], who proposed that the complexity of a waveform may be represented by what Mandelbrot [17] named {${\textit{fractal dimension}}$}, (represented as {${\Phi}$} in this communication). For this purpose Katz [15] reported that the fractal dimension might be measured empirically by sampling the waveform at {$\textit{N}$} points evenly spaced on the abscissa. This procedure discretizes the waveform into {${N'=N-1}$} segments and then, according to Katz's equation (5):}

		\begin{equation*}
			\Phi =\frac{\log (N')}{\log (N')+\log (d/L)}
		\end{equation*}
		
		where \textit{d} is the \textit{planar extent} of the curve [17, Chapter 12] and \textit{L} is the length of the curve, both of them defined as:
		
	\begin{equation}
		\begin{gathered}
			d=\text{max } [\text{dist}(i,j)] 
			\\ L= \overset{N'}{\underset{i=0} \sum} {\text{dist}(i,i+1)}
		\end{gathered}
	\end{equation}
		
		where {\textquotedbl}max{\textquotedbl} stands for the maximum ${\text{dist}(i,j)}$, the distance between points \textit{i} and \textit{j} of the curve. For curves that do not cross themselves usually, but not always, ${d=\text{max dist(1,i)}}$. Yet, as I prove in the Results section of this communication, ${D_k}$ does not measure $\Phi$. Here I describe an algorithm to calculate {\textit{D}, }an empirical approximation to $\Phi$, which is easy to set up on a computer, fast to calculate and one that lacks the shortcoming of the Katz's [15] equation.

	\section[2 Methods]{2 Methods}
		
		\subsection[2.1 Computer implementation of the algorithms]{2.1 Computer implementation of the algorithms}
			
			All calculations were programmed in C++ (IBM C Set++, IBM Boca Raton, FL) under OS/2{${^{TM}}$} Warp 3.0 (IBM Boca Raton, FL) on a 90 MHz Pentium{${^{TM}}$} (Intel Corporation, Portland, OR) computer. Normally distributed (Gaussian) random variables with mean zero and variance one, \textit{N(0,1)}, were generated using an optimized Box and Muller algorithm [3, 22]. The algorithm used to generate random variables [named here \textit{U(0,1)}]
			distributed with equal probability in the interval \textit{(0,1)} was a \ C++ implementation of the method of Kirkpatrick and Stoll [16]. This implementation was not only very fast, but was very uniformly random and had a very long cycle length. Under different operating systems such as MS-DOS 6.21 (Microsoft Corp., Redmont, WA), OS/2 Warp 3.0 (IBM Corp., Boca Raton, FL) and SunOS 4.1.3 (Sun Microsystems Inc., Mountain View, CA) program could produce between 1 and 8 billion uniformly distributed random numbers without repetitions. The algorithm may be downloaded via the Internet
			using anonymous ftp from toxico.ivic.gob.ve as pub/os2/random.zip. Other programs used in this communication may be also downloaded from this anonymous ftp server at /pub/complexity.
			
		\subsection[2.2 A Simple Method to Calculate Fractal Dimension of Waveforms]{2.2 A Simple Method to Calculate Fractal Dimension of Waveforms}
		
			An expression to calculate the fractal dimension of a waveform is obtained starting from the definition of Hausdorff dimension ($D_H$). Mandelbrot's definition calls \textit{fractal} (see for example [18]) to a set whose Hausdorff dimension is not an integer. The Hausdorff dimension [17] of a set in a metric space (see Barnsley [2] for a very readable discussion on metric spaces) may be expressed as:

			\begin{equation}
				D_{h}=\underset{\epsilon \rightarrow 0}{\lim }{\frac{-\ln [N(\epsilon )]}{\ln (\epsilon )}}
			\end{equation}
			
	{\selectlanguage{english}		
where \textit{N(${\epsilon}$)} is the number of open balls of a radius \textit{${\epsilon}$} needed to cover the set. In a metric space, given any point \textit{P}, an open ball of center \textit{P} and radius ${\epsilon}$, is a set of all points \textit{x} for which \text{dist}(P,x) < ${\epsilon}$. A line of length \textit{L} may be divided into ${N(\epsilon)=L/(2 \cdot \epsilon)}$ segments of length ${2 \cdot \epsilon}$, and may be covered by \textit{N} open balls of radius ${\epsilon}$. Thus, equation (2) may be rewritten as }
			
			{\selectlanguage{english}
				\begin{equation}
					D_{h}=\underset{\epsilon \rightarrow 0}{\lim }\left[\frac{-\ln (L)+\ln (2\cdot \epsilon )}{\ln (\epsilon )}\right]=\underset{\epsilon \rightarrow 0}{\lim }\left[1-\frac{\ln (L)-\ln (2)}{\ln (\epsilon )}\right]=\underset{\epsilon \rightarrow 0}{\lim }\left[1-\frac{\ln (L)}{\ln (\epsilon )}\right]
				\end{equation}
			}
			
			Waveforms are planar curves in a space with coordinates usually having different units. Since the topology of a metric space does not change under linear transformation, it is convenient linearly to transform a waveform into another in a normalized space, where all axes are equal. I propose to use two linear transformations that map the original waveform into another embedded in an equivalent metric space. The first transformation, normalizes every point in the abscissa as:
			
			\bigskip
			
			\begin{equation}
				x_{i}^{\text{*}}=\frac{x_{i}}{x_{\mathit{max}}}
			\end{equation}

			Where \textit{x\textsubscript{i}} are the original values of the abscissa, and \textit{x\textit{\textsubscript{max}}}
			is the maximum ${x_i}$. The second transformation normalizes the ordinate as follows:
			
			\begin{equation}
				y_{i}^{\text{*}}=\frac{y_{i}-y_{\mathit{min}}}{y_{\mathit{max}}-y_{\mathit{min}}}
			\end{equation}
			
			where ${y_{i}}$ are the original values of the ordinate, and ${y_{min}}$ and ${y_{max}}$ are the minimum and maximum ${y_{i}}$, respectively. These two linear transformations map the \textit{N} points of the waveform into another that belongs to a unit square. This unit square may be visualized as covered by a grid of ${N \cdot N}$ cells. \textit{N} of them containing one point of the transformed waveform. Calculating {\textit{L}} of the transformed waveform and taking ${\epsilon=1/(2 \cdot N')}$ equation (3) becomes

			\begin{equation}
				D_{h}=\Phi \approx D=1+\frac{\ln (L)}{\ln (2 \cdot N')}
			\end{equation}
			
			The approximation to ${\Phi}$ expressed in equation (6), improves as ${N' \rightarrow \infty}$.
			
		\subsection[2.3 An Approximate Expression for the Variance of \textit{D}.]{2.3 An Approximate Expression for the Variance of \textit{D}.}
		
		Although ${\Phi}$ is a topological invariant of a set or a metric space, \textit{D} is only an empirical estimate of $\Phi$ with some uncertainty based on a set of points sampled from a waveform; \textit{D} is thus a random variable. The relationship between ${\Phi}$ and \textit{D}, is similar to the one between the mean of a \textit{population} ($\mu$) and the mean ${\bar{{x}}}$ estimated sampling a subset of the population; although ${\mu}$ is an invariant for the population, ${\bar{{x}}}$  will change with sampling. Just as ${\bar{{x}}}$ converges to ${\mu}$ as the sample size approaches the size of the population, \textit{D} converges to ${\Phi}$ as ${N'\rightarrow \infty}$. I will now derive an expression to ${\text{var}(D)}$ (the variance of \textit{D}) for the estimates of {\textit{D}} obtained by sampling ${N{\textquotesingle}}$ points from a waveform. It should be obvious from the derivation of ${\text{var}(D)}$ and the non stationary character of the values of \textit{D} determined with equation (6), that ${\text{var}(D)}$ does not provide information on the asymptotic value of \textit{D} obtained as ${N'\rightarrow \infty}$. The variance of \textit{ D} may be estimated starting from the following expression:
		
		\begin{equation}
			\text{var}(D)=\text{var}\left[1+\frac{\ln (L)}{\ln (2\cdot N\text{{\textquotesingle}})}\right] = \text{var}\left[\frac{\ln (L)}{\ln (2\cdot N\text{{\textquotesingle}})}\right]\hfill 
		\end{equation}
		
		The approximate solution to equation (7) may be obtained recalling that the variance of any function of ${x_i}$ independent random variables is obtained with a Taylor series (see for example [5]) as:

		\begin{equation}
			\text{var}\left[f\left(x_{1},x_{2},\ldots ,x_{i},\ldots ,x_{k}\right)\right]\approx \overset{k}{\underset{i=1}{\sum }}{\left[\left(\frac{\partial \left[f\left(x_{1,}x_{2},\ldots ,x_{i},\ldots ,x_{k}\right)\right]}{\partial x_{i}}\right)^{2}\cdot \text{var}(x_{i})\right]}\hfill
		\end{equation}

		which for equation (7) produces:

		\begin{equation}
			\text{var}(D)=\frac{\text{var}(L)}{L^{2}\cdot \text{ln}(2\cdot N\text{{\textquotesingle}})^{2}}\hfill 
		\end{equation}
		
		since \textit{L} is the sum of \textit{N'} segments of length $\Delta y$, equation (9) is equivalent to
		
		\begin{equation} 
			\text{var}(D)=\frac{N\text{{\textquotesingle}}\cdot \text{var}(\Delta y)}{L^{2}\cdot \ln (2\cdot N\acute{})^{2}} \hfill 
		\end{equation}
		
		where ${\text{var}(\Delta y)}$ may be estimated from the data as:

		\begin{equation}
			\text{var}(\mathit{{\Delta}y})=\overset{N}{\underset{i=1}{\sum }}{\frac{\left(\Delta y_{i}-{\overline{{\Delta y}}}\right)^{2}}{N\text{{\textquotesingle}}}}\hfill
		\end{equation}

		where  ${\overline{{\Delta y}}}$ is the mean segment length.
		
	\section[3 Results]{3 Results}
	
		\bigskip
		
		\subsection[3.1 Predictions with Katz's expression for ${D_{k}}$]{{\textrm{\textup{3.1 Predictions with Katz's expression for \textit{D}${_k}$}}}}
		
			\bigskip
			
			It is possible to show \textit{for any waveform} that, if the ratio ${d/L}$ is constant, equation (5) of Katz [15] has the following limit

			\begin{equation}
				{\selectlanguage{english}
				\underset{N' \rightarrow \infty }{\lim } \left[D_{k}=\frac{\log (N')}{\log (N')+\log (d/L)}\right]=1
				}
			\end{equation}

			\bigskip

			This result means that ${D\textsubscript{k}}$ does not measure ${\Phi}$  of a waveform. Equation (12) shows that, according to Katz's expression, as we obtain more information on any waveform we will find that \textit{all of them} are straight lines with ${D\textsubscript{k} = 1}$. This holds true for all waveforms for which ${d/L}$ is asymptotically constant after sampling ${m < \infty}$ points. For ${N' < \infty}$ equation (12) also implies that the value of $D_k$ is determined by the (arbitrary) choice of $N$. The condition of the limit, ${N'\rightarrow \infty}$, may be fulfilled if ${t_{max}}$ (the sampling duration) stays constant but the sampling interval ${\delta \rightarrow 0}$.

			\begin{figure}
				\centering
				\includegraphics[width=10cm]{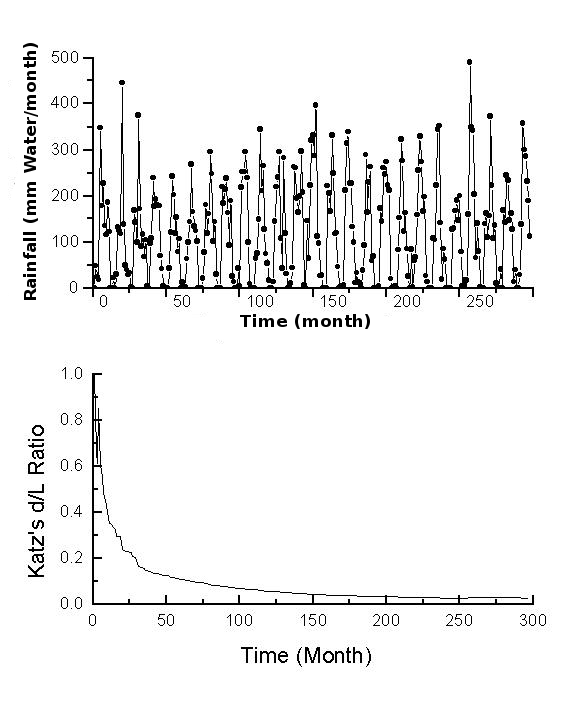}
				\caption{\textbf{Properties of the ratio ${d/L}$ for a rainfall plot.} The top of the figure presents data on rainfall (mm water/month) in the central plains of Venezuela. The lower part of the figure is the ratio of planar extent (\textit{d}) divided by the length of the curve (\textit{L}). Other details in the text of the communication.}
			\end{figure}

			The (asymptotic) constancy of the ratio ${d/L}$ may be verified experimentally for many functions, even if \textit{m} is small. Figure 1 contains data on rainfall in a region of the central plains of Venezuela (Turagua Ranch, Apure State, 68{${^\text{o}}$} 19{\textquotesingle} 27{\textquotedbl} W, 7{${^\text{o}}$} 48{\textquotesingle} 15{\textquotedbl} N, data are monthly totals measured from January 1969 to May 1994). The data were chosen as an example of a waveform occurring in nature. The figure shows that \textit{d/L} becomes constant for values of ${N'> 150}$. Since some periodicity is evident in the rainfall curve, I show results for a waveform characteristic of \textit{white noise} in figure 2. Is customary to call noise to a variety of randomly fluctuating waveforms; white noise, is a random function that has a constant power spectrum at any frequency. The 300 points in the figure are uniformly distributed random events in the interval \textit{(0,1)}, joined by straight lines to create a waveform. As shown in the lower part of figure 2, even for a random function the ratio \textit{d/L} approaches constancy for ${N' > 150}$. The data in these two figures suffices to prove that for many functions the limit expressed by equation (12) holds true. It is clear, thus, that ${D_k}$ does not measure ${\Phi}$ of waveforms as claimed by Katz.

			\begin{figure}
				\centering
				\includegraphics[width=10cm]{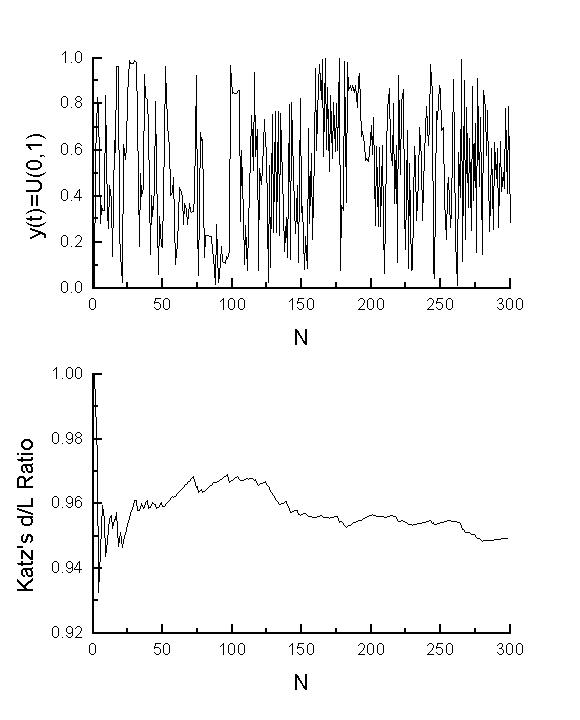}
				\caption{\textbf{Properties of the ratio \textit{d/L} for a random curve.} The top of the figure presents a waveform built by joining with straight lines a series of pseudo-random uniform points in (0,1). The lower part of the figure is the ratio \textit{d/L}. Other details as in figure 1 and in the text of the communication.}
			\end{figure}

		\subsection[3.2 Efficiency of Equation (6) to Estimate Fractal Dimension of Several Types of Random \ Signals ]{3.2 Efficiency of Equation (6) to Estimate Fractal Dimension of Several Types of Random Signals }
		
			To test the efficiency of \textit{D} for estimating the fractal dimension of waveforms, I selected Brownian and white noises. A continuous\nobreakdash-time random walk or Brownian process (or Brownian noise) is defined, for a time step $\Delta t \rightarrow 0$, as
			
			\begin{equation}
				y(t+\Delta t)=y(t)+\Delta y(t)
			\end{equation}
			
			where \textit{${\Delta}$y(t)} is Gaussian, of mean zero and variance proportional to \textit{${\Delta}$}\textit{t} [25]. A Brownian process is known to have a fractal dimension equal to 1.5 [12]. Thus it seems ideal to test the ability of equation (6) to predict {\textit{D}}. Calculating fractal dimension of a planar curve with the Hurst's exponent is possible (\textit{H}) [13, 14] with the following relation [12]:
			
			\begin{equation}
				D=2-H
			\end{equation}
			
			Thus, 20 sets of 100000 points with $y(0) = 0$ were generated and $\Delta y(t)=N(0,1)$ . The value of {\textit{D}} was calculated for each of them with equation (6), and from values of \textit{H} estimated as indicated by Hastings and Sugihara [12]. Besides the Brownian noise represented by equation (13), other three types of signals were used as examples. In one of them, the variable \textit{N(0,1)} in equation (12) was replaced by \textit{U(\nobreakdash-0.5,0.5)} (Non Gauss Walk, in Tables 1 and 2).
			
			\bigskip
			
			{\centering
			\textbf{TABLE 1}
			\par}
			
			{\centering
			\textbf{Numbers of Runs Above and Below the Median of Several Waveforms}
			\par}
			
			\begin{center}
				\begin{longtable}{ll}
					\hline
					\textbf{\textit{Type of Waveform}} & {\centering \textbf{Number of Runs}} \\
					\textit{\textit{Brownian Noise}} & {\centering 405 ${\pm}$ 53} \\
					\textit{\textit{Non Gauss Walk}} & {\centering 258 ${\pm}$ 24} \\
					\textit{\textit{Gaussian White Noise}} &  {\centering 49992 ${\pm}$ 34} \\
					\textit{\textit{Uniform White Noise}} &  {\centering 49977 ${\pm}$ 20} \\
					\hline
				\end{longtable}
				\begin{quote}
				\begin{footnotesize}
					All values are means ${\pm}$ standard error of mean calculated for 20 sets of traces consisting of 100000 points/trace.
				\end{footnotesize}
				\end{quote}
			\end{center}
			
			\bigskip
			
			In the other two cases \textit{y(t)} equals to \textit{ U(\nobreakdash-0.5,0.5)} or to \textit{N(0,1)} (Uniform White Noise and Gaussian White Noise, respectively, in tables 1 and 2). These procedures produce signals with different statistical properties. The data in table 1 summarizes a study of runs above and below the median [25, pp 144 \nobreakdash- 150] of the four kinds of traces described. The points composing all the traces called here uniform and Gaussian white noises, produced a number of runs $\approx$50,000 as it should be if they are distributed randomly about the median [25, pp 144\nobreakdash-150]. The number of runs about the median of Brownian and non Gaussian walks are not only distinct from the white uniform and Gaussian white noises, but distinct between themselves (${P = 2 \cdot 10^{-5}}$, Student's t test).
			
			\newpage
			
			\begin{center}
			\textbf{TABLE 2}
 			\end{center}
			
			\begin{center}
			\textbf{Values of Fractal Dimension of Waveforms Calculated from of Hurst's Exponents, Equation (5) of Katz (1988) (\textit{D${_k}$}) and with Equation (6) of this paper (\textit{D})}
			\end{center}
			\begin{center}
				\begin{longtable}[c]{p{1.3in}p{0.76in}p{0.76in}p{0.76in}p{0.76in}p{0.76in}p{0.76in}}
					\hline
					&\underline{\begin{small}\textbf{Methods Based on Hurst's Exponent Determination}\end{small}}   \\ 
					\textbf{Type of Waveform} & {2\textsuperscript{nd} Moment} & {Growth of Range} & {Local Growth} & {Power Spectrum} & {\textbf{\textit{D\textsubscript{k}}}} & {\textbf{\textit{D}}} \\
					\hline
					\textbf{\textit{Brownian}} &1.4988 ${\pm}$0.0013 & 1.3578 ${\pm}$0.0007 & 1.4993 ${\pm}$0.0009 & 1.5005 ${\pm}$0.0032 & 1.0270 ${\pm}$1.7${\cdot 10^{-5}}$ & 1.4261 ${\pm}$0.0051 1.4359 ${\pm}$0.0037\textsuperscript{*} \\ 
					\textbf{\textit{Non Gauss}} & 1.4978 ${\pm}$0.0016 & 1.3688 ${\pm}$0.0009 & 1.4979 ${\pm}$0.0015 & 1.5000 ${\pm}$0.0003 & 1.0034 ${\pm}$1.6${\cdot10^{-6}}$ & 1.4166 ${\pm}$0.0077 1.4332 ${\pm}$0.0007\textsuperscript{*} \\ 
					\textbf{\textit{Gaussian}} & 1.9999 ${\pm}$0.0001 & 1.7341 ${\pm}$0.0004 & 2.0000 ${\pm}$0.0001 & 1.5000 ${\pm <10^{-4}}$ & 1.0428 ${\pm}$1.8${\cdot 10^{-5}}$ & 1.7763 ${\pm}$0.0009 1.8043 ${\pm}$0.0007\textsuperscript{*} \\ 
					\textbf{\textit{Uniform}} & 1.9999 ${\pm}$0.0001 & 1.8596 ${\pm}$0.0001 & 2.0003 ${\pm}$0.0001 & 1.5000 ${\pm <10^{-5}}$ & 1.0065 ${\pm}$4.9${\cdot 10^{-6}}$ & 1.8531 ${\pm<10^{-4}}$ 1.8765 ${\pm<10^{-4*}}$\\
				\hline
				\end{longtable}

				\begin{footnotesize}
					\begin{flushleft}
						All values are means ${\pm}$ standard error of mean calculated for 20 sets of traces consisting of 100000 points/trace, except for data marked with asterisks calculated for 20 sets of 1 million points.	
					\end{flushleft}
				\end{footnotesize}
			\end{center}

			It is apparent from the data in table 2 that (unless calculating \textit{H} from the growth of range) the estimates of ${\Phi}$ based on Hurst{\textquotesingle}s exponents were close to 1.5, the fractal dimension of Brownian noise. This achievement is, however, overshadowed by the lack of ability of the methods based on Hurst's exponents to differentiate signals with distinct statistical properties. The only procedure based on \textit{H} that predicted different values of ${\Phi}$ for the four types of signals was based on growth of range (${P{\ll} 10^{-6}}$). Yet, this procedure grossly underestimated ${\Phi}$ of Brownian noise. The procedures based on determining \textit{H} from the 2{\textsuperscript{nd}} moment or from the local growth, probably overestimate the real value of \textit{${\Phi}$}. This may be concluded from the values of \textit{${\Phi}$} calculated for uniform and Gaussian white noises that do not differ from 2. In contrast with these results, the values of \textit{D}, for the different types of curve tested (Table 2), are statistically distinct (${P{\ll}10^{-4}}$). Equation (5) of Katz (1988) was also applied to each of the twenty traces of the four kinds of noise mentioned above; as shown in table 2 the value of ${D^{k}}$ was very close to 1 for all the four types of noise studied.

			\bigskip
		
		\subsection[3.3  Convergence of D towards a steady state]{3.3 
		Convergence of \textit{D} towards a steady state}
		
		\bigskip

		The data presented in figures 1 and 2 were also used to show that the values of \textit{D} and its standard deviation converge relatively quickly to a steady state. This is illustrated in figure 3.
		
		\begin{figure}
			\centering
			\includegraphics[width=10cm]{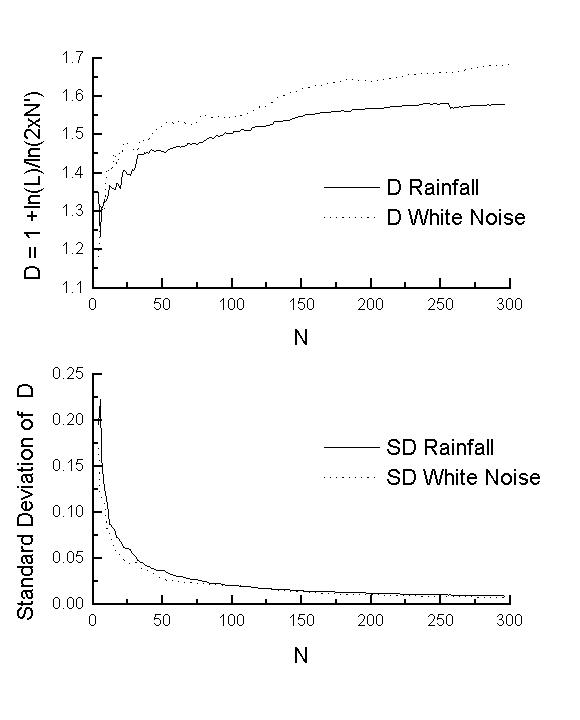}
			\caption{\textbf{Fractal dimension (\textit{D}) and standard deviation of \textit{D} for the curves in figures 1 and 2.} The top of the figure presents the values of fractal dimension for the rainfall data in figure 1 calculated with equation (6). The standard deviations of \textit{D} calculated with equation (10) for both curves are in the lower part of the figure. Thick lines correspond to rainfall data, and thin lines to the random curve in figure 2.}
		\end{figure}

		\bigskip
		
		\subsection[The asymptotic convergence of \textit{D} to $\Phi$. An analytical solution for the Koch's triadic curve]{The asymptotic convergence of \textit{D}  to $\Phi$. An analytical solution for the Koch's triadic curve}
		
		\bigskip
		
		Although I derived equation (6) for waveforms, i.e. planar curves that are sets of pairs of points $(x_i, y_i)$ such as $x_i \rightarrow \infty$ as $i \rightarrow \infty$. Yet at least in some instances the usefulness of \textit{D} to estimate $\Phi$ extends beyond the realm of waveforms, I will show this using the famous Koch triadic curve shown in figure 4.
		
		\begin{figure}
			\centering
			\includegraphics[width=7cm]{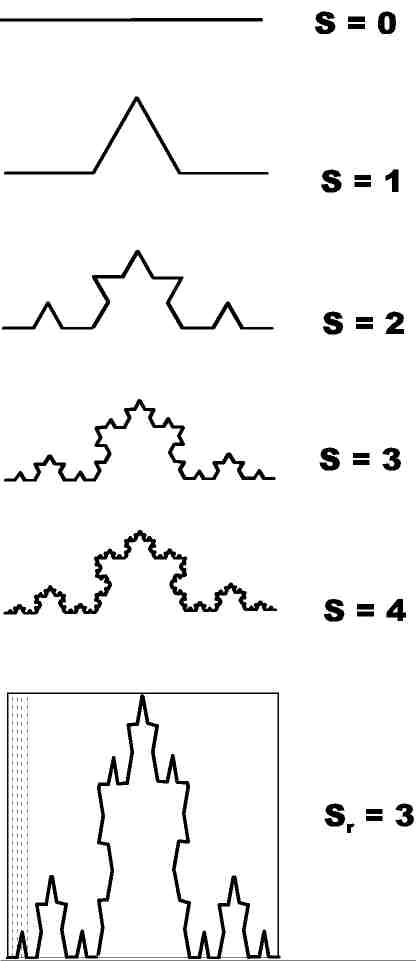}
			\caption{\textbf{Construction of the triadic Koch curve}. The triadic Koch curve is constructed as the limit of a 	sequence of simple iterative steps. Starting with a line of unit length as shown at the top of the figure (stage 0 or s = 0, in the figure), then at each stage the middle third of each line of length ${\lambda}$ is replaced by to equal segments of length ${\lambda}$/3 forming to sides of an equilateral triangle. Proceeding for an infinite number of steps, one obtains the triadic Koch curve. At the bottom of the figure stage 3 appears rescaled into a unit square, see the text for other details.}
		\end{figure}
		
		It may be easily verified that the following properties hold for the triadic curve at any step \textit{S}:

		\begin{equation*}
			\begin{gathered}
				n_{s}=2^{2 \cdot S}\hfill
				\\L=\left[\frac{4}{3}\right]^{S}\hfill \\l_{s}=3^{-S}\hfill
				\\n_{\mathit{hs}}=\frac{2^{2\cdot S}-1}{3}+1\hfill
				\\n_{\mathit{is}}=2^{2\cdot S}-\left(\frac{2^{2\cdot
				S}-1}{3}+1\right)\hfill \\n_{v}=2^{2\cdot S}-1\hfill
				\\K_{h}=\frac{1}{\sqrt{12}}\hfill 
			\end{gathered}
		\end{equation*}
		
		Where \textit{S} is the stage number (as used in figure 4), $n_s$ is the number of segments forming the curve, \textit{L} is the curve's length, $l_s$ is the length of each segment, ${n_{hs}}$ is the number of segments that are {\textquotedbl}horizontal{\textquotedbl} segments (i.e. may be extended as parallels to the line in stage 0) ${n_{is}}$ is the number of {\textquotedbl}inclined{\textquotedbl} (i.e. are not horizontal as defined above) and ${n_v}$ is the number of vertexes in the curve ${K_h}$ is the altitude of the equilateral triangle built in stage 1, measured perpendicularly to the line that prolongs the two horizontal segments at its base. Thus if one centers open balls of radius
		
		\begin{equation*}
			\epsilon =\frac{3^{-S}}{2}
		\end{equation*}
		
		on both ends of the triadic curve, and on every intersection of segments in the curve we have that
		
		\begin{equation*}
			N(\epsilon )=1+2^{2 \cdot S}
		\end{equation*}
		
		of such open balls are required to cover the curve. From the expressions for ${\epsilon}$ and $N(\epsilon)$ and equation (2) we get the fractal (Hausdorff) dimension as:
		
		\begin{equation*}
			{\selectlanguage{english}
			D_{h}=\underset{S\rightarrow \infty }{\lim }-{\frac{\ln\left(1+2^{{2S}}\right)}{\ln \left(\frac{3 ^{-S} }{2} \right)}}=\frac{\ln(4)}{\ln (3)}=1.2618\ldots }
		\end{equation*}

		Which is also the similarity and covering dimensions of the triadic curve.
		
		To test the ability of equation (6) for predicting ${D_H}$ in the case of the triadic curve we have to transform the curve as follows:
		
		\begin{equation*}
			\begin{gathered}x_{i}^{\text{*}}=x_{i}\hfill\\y_{i}^{\text{*}}=y_{i}\cdot \sqrt{12}\hfill \end{gathered}
		\end{equation*}
		
		This is shown at the bottom of figure 4 for the 3{\textsuperscript{rd}} stage of the generation process. The transformation does not modify the length of the horizontal components of the Koch curve \ but extends the length of all inclined segments which becomes

		\begin{equation*}
			l_{\mathit{is}}=\sqrt{\frac{1}{6^{2\cdot S}}+\frac{12}{6^{2 \cdot S}}}=\frac{\sqrt{13}}{6^{S}}
		\end{equation*}
		
		And then for the curve at stage \textit{S} we have
		
		\begin{equation*}
			L=n_{\mathit{hs}} \cdot l_{s}+n_{\mathit{is}} \cdot l_{\mathit{is}}=\left(\frac{2^{2 \cdot S}-1}{3}+1\right)\cdot{\frac{1}{3^{S}}}+\left[2^{2 \cdot S}-\left(\frac{2^{2 \cdot S}-1}{3}+1\right)\right] \cdot {\frac{\sqrt{13}}{6^{S}}}
		\end{equation*}
		
		then, in order to provide that every vertex of the curve corresponds to one cell of the normalized square we have to let

		\begin{equation*}
			N'=3^{S}
		\end{equation*}
		
		which is sketched as dotted lines in figure 4 ($s_r = 3$). Then by replacing in equation (6)

		\begin{equation*}
			{\selectlanguage{english}
			D=\underset{S\rightarrow \infty }{\lim }\left[1+\frac{\ln \left\{\left(\frac{2^{2\cdot S}-1}{3}+1\right)\cdot {\frac{1}{3^{{S}}}}+\left[2^{2 \cdot S}-\left(\frac{2^{2 \cdot S}-1}{3}+1\right)\right]\cdot {\frac{\sqrt{13}}{6^{S}}}\right\}}{\ln \left(2\cdot 3^{S}\right)}\right]=\frac{\ln (4)}{\ln (3)}=1.2618\ldots }
		\end{equation*}
		
		as it should be.
		
		\subsection[3.5 An Example of Application to a Non-Stationary Process: The Spanish Flu Epidemics of 1918]{3.5 An Example of Application to a Non-Stationary Process: The Spanish Flu Epidemics of 1918}
		
		\bigskip
		
		In 1918 an epidemic influenza started in Spain [7] raided the world and reached Caracas {\textquotedblleft}One case, two, three, six, a hundred, on the Capitol it fell like fog, the raid swept mercilessly from outskirts to center{\textquotedblright} (my translation of a passage by Pocaterra [21]). The top of figure 4 presents a plot of casualties/day due to Spanish flu in the Federal District (Venezuela) during October and November, 1918; in 1920, the Federal District had 140,132 inhabitants [4]. \ The data on the epidemics is included here as an example of a waveform without apparent periodicity [redrawn from measurements of a graphic in D\'avila [7].
		
		\begin{figure}
			\centering
			\includegraphics[width=10cm]{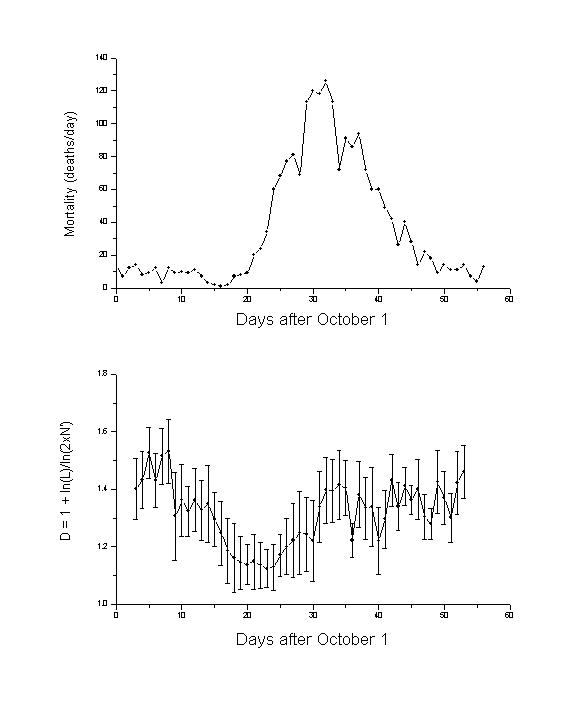}
			\caption{\textbf{Fractal analysis of the Spanish flu epidemic of 1918 in the Federal District, Venezuela.} The top of the figure is a plot of casualties/day due to the Spanish flu in the Federal District during October and November 1918. The lower part of the figure is a plot of fractal dimension and its standard deviation, of the mortality curve above. The fractal dimension (solid line) was calculated using equation (6) of this communication. Vertical lines delimit an interval extending one standard deviation [calculated with equation (10)] above and below the fractal dimension predicted by equation (6). The values were calculated for a window of 7 points, which was displaced by one day each time. Data are plotted at the point corresponding to the center (4{\textsuperscript{th}} day) of the window. See text for other details.}
		\end{figure}	
		
		\bigskip
		
		The procedure to estimate fractal dimension of a waveform, and its standard deviation expressed by equations (6) and (10) is very fast and can be applied to small amounts of data. The lower part of figure 5 is a plot of \textit{D} calculated from a sliding window of seven points of casualties/day by Spanish flu (solid line). For this purpose fractal dimension and its standard deviation (dotted lines with triangles) were estimated from sets of seven points starting with the datum corresponding to October 1. After each estimate, the window was moved one day forward, the calculation repeated, and so on. The data in the figure is presented on top of the 4{\textsuperscript{th}} day (the middle) of the window. The fractal analysis shows that in spite of the small number of points used, it is possible to prove that the course of the epidemics exhibited significant changes in fractal dimension. Initially, it behaved like a Brownian process with \textit{D} between 1.4 and 1.5. Then {\textit{D}}{ dropped, first to 1.3 for approximately 1 week. \ Then again to 1.1 during the onset of the burst that occurred between October 16 and November 15. Finally, the fractal dimension raised to 1.3 and became highly variable, this coincided with the most drearisome part of the epidemics and the few weeks that followed. The wide interval estimator in figure 5 (delimited in by dotted lines with triangles) is one sample standard deviation of \textit{D}. Since for a particular waveform \textit{N} may be taken as constant, the variance of the \textit{mean D} [the square of the standard error of mean (SEM)] is
		
		\begin{equation*}
			\text{var}(\overline{{D}})=\frac{\text{var}(D)}{N\text{{\textquotesingle}}}=\frac{\text{var}(\Delta y)}{\left[L\cdot \ln (2\cdot N\text{{\textquotesingle}})\right]^{2}}\hfill
		\end{equation*}
				
		For the sliding window in figure 5, $N' = 6$, thus  $\text{var}(\overline{{D}})=\text{var}(D)/6 $, and since
		
		\begin{equation*}
			\text{SEM}=\sqrt{\text{var}(\overline{{D}})}
		\end{equation*}

		this means that the SEM is 2.45 times smaller that the standard deviation shown in figure 5.

	\section[4. Discussion]{4. Discussion}
		
		At the core of the analysis described in this communication is a Monte Carlo random process simulation. It may be described as follows [6]: 
		
		\begin{quote}
			In most of the applications of probability theory one makes a mathematical formulation of a stochastic problem (i.e., a problem where chance plays some part), and then solves the problem by using analytical or numerical methods. In the Monte Carlo method, one does the opposite: a mathematical problem is given, and one constructs a game of chance that in some way leads to the given problem.
		\end{quote}

		\bigskip
		
		In the problem we are dealing with, the game of chance is the process to construct random curves with a known fractal dimension (as it is the case of the Brownian process). The problem is to detect the ability of different procedures described in the literature to find the fractal dimension of such waveforms. In particular my interest was centered in analyzing the efficiency of equation (6) and of the method described by Katz [15] as useful to find the fractal dimension of waveforms. Also, I have applied equation (6) to one waveform describing rainfall and to another describing the mortality due to Spanish flu in Caracas in 1918. \textit{The two latter curves were used to illustrate the feasibility of approximating the fractal dimension of waveforms arising in Nature using only the few points that are available. This is not intended as an apology for fractal analysis in epidemiology or meteorology.} Yet, many authors have used this analysis. Examples of the use of the fractal dimension in meteorology are found in Mandelbrot [17], Hastings and Sugihara ([12, Chapter 11), Schroeder ([23, Chapter 5) or Tsonis ([24], Chapter 10). Examples of the use of the fractal analysis in epidemiology also exist [23; 24].
		
		In this paper I propose the method expressed by equation (6). The method is shown to be at least as fast as the
		method of Katz [15]. The method is accurate using few experimental points and is free of the shortcomings of the latter, and is very simple to set up in a computer. An approximate expression for the variance of the fractal dimension estimate is also provided. The speed of the algorithm may be appreciated by considering its performance on a slow computer such as a 40 MHz 80386 PC (under OS/2 Warp 3.0 and with the IBM C Set++ compiler); the calculation of {\textit{D}} for a white noise trace consisting of 100000 points took 7 {\textit{s}}. Under similar conditions, my implementations in C++ of the programs in the appendix of Hastings and Sugihara [12] to calculate H with the local growth algorithm also took 7 {\textit{s}}. Yet, calculating H (as indicated by Hastings and Sugihara [12] but programming in C++) with the growth of range and growth of 2{\textsuperscript{nd}} moment algorithms took 3042 {\textit{s}}. An algorithm based on estimating fractal dimension from the correlation dimension [9] for the  same waveform and machine, in my experience, takes several days of computer time.
		
		Equation (6) is not only easy to set up and fast to calculate on a computer, but also predicts satisfactorily the value of ${\Phi}$. As shown by the data in table 2, equation (6) produces values of {\textit{D}} that are closest to the value ${\Phi}$ for Brownian for Brownian noise, which also reflect the differences existing among the four types of curves studied. Yet, the data in table 2 shows that the values of {\textit{D}} are lower than 1.5, the known value of fractal dimension of a Brownian process [12, 17]. This underestimation is probably partly due to my procedure to generate the Brownian waveforms; equation (13) describes a Brownian process in the limit ${\Delta t \rightarrow 0}$, which is not so in this paper where ${\Delta t}$ was always 1 and ${y( \Delta t ) = N(0,1)}$. Under this conditions the waveform is composed from straight line segments of ${\Phi = 1}$, which were longer in comparison with the length of the curve than it would have been if ${\Delta t \rightarrow 0}$. Thus ${\Phi}$ of the \textit{approximate} Brownian waveforms used must be {\textless}1.5 as suggested by the values of {\textit{D}}, which is thus, probably more accurate than it seems in table 2. Should this be the case, the values of ${\Phi}$ calculated using procedures based on Hurst{\textquotesingle}s exponent that are closer to 1.5, are larger that the \textit{real} value for the \textit{simulated} curves. This overestimation of \textup{${\Phi}$} by some procedures based on \textit{H}, is probably why they produced ${\Phi=2}$ for the white noise waveforms in table 2. A value of ${\Phi=2}$ for a planar curve that does not cross itself implies that it
		completely covers the plane, this does not occur for any of the waveforms studied here. 
		
		Another characteristic of equation (6) is that it produces approximate values of the fractal dimension which are useful even when a few points are considered. This is shown by the analysis of data from the Spanish flu epidemics in figure 4. In the epidemics studied, the values of fractal dimension clearly differentiate several stages of this process. The differences are clearer from \textit{ D} values that from the bare observation of the mortality curve, especially in regards with the outburst. The outburst, is singled by the fractal analysis not just as an increase if frequency of casualties, but as a stage where the system changes its properties. The passage of Pocaterra [21] cited above, suggest that the change may be related to the spatial distribution of the casualties. In connection with this example of use \textit{D} with small values of \textit{N}\textquotesingle, it is good to notice that the intermediate form of equation (3):
		\begin{equation*} 
		D=1 + \frac{\text{ln}(L)-\text{ln}(2)}{\text{ln}(2\cdot N\text{{\textquotesingle)}}} \text{\qquad \qquad \qquad (6a)}
		\end{equation*}
		
is more adequate for small \textit{N}\textquotesingle.

		Some words of caution are in order at this point. First, according to Mandelbrot ([17], pg. 15 and Ch. 39):
		
		\begin{quote}
			A fractal is by definition a set for which the Hausdorff\nobreakdash-Besicovitch dimension strictly exceeds the topological dimension. Every set with a non-integer \textit{D} is a fractal.
		\end{quote} 
		
		Thus, any planar curve with ${1 < D_H < 2}$ is \textit{fractal}. However, associated to the notion of fractal are often notions such as \textit{self\nobreakdash-similarity}, or \textit{self\nobreakdash-affinity} ([17], pp. 349 and 350), which are not granted by just observing that for a certain waveform ${1 < D_H < 2}$. Secondly, be careful when using ${\Phi}$ to estimate \textit{complexity}, since a larger ${\Phi}$ does not necessarily mean more \textit{complexity}; the reader is advised to check the discussion of Mandelbrot ([17], pg. 41) on this subject. Thirdly, in the deduction of equation (10), the variance of \textit{D}, it is implicitly assumed that ${\Delta y}$ is \textit{homoscedastic} (has the same variance) in the range of the abscissa where \textit{D} is calculated; homoscedasticity does not always exists, and equation (10) may not be valid under those circumstances.
		
		This paper is a piece of applied mathematics. It is intended mainly to aid in the analysis of waveforms, it stemmed from my need of interpreting waveforms recorded from nervous tissue and from finding that Katz{\textquotesingle}s [15] work is useless. Here I present a fast algorithm to calculate the fractal dimension of waveforms. It is shown experimentally that as ${N' \rightarrow \infty}$ the value of \textit{D} predicted by equation (6) converges to the fractal dimension of a Brownian noise, a random fractal set. Also, it is demonstrated that as ${N' \rightarrow \infty}$, \textit{D} converges to the fractal dimension, ${D_H}$, in the case of a triadic Koch{\textquotesingle}s curve, a non-random fractal set. These results are quite natural, since equation (6) is nothing but an asymptotic approximation to ${D_H}$, the Hausdorff dimension, applied to waveforms linearly transformed into a unit square.
		
		I have presented results that suggest ${D \rightarrow D_H }$ as ${N' \rightarrow \infty}$ for {self\nobreakdash-affine} Brown noises and demonstrated this for the {self\nobreakdash-similar} triadic Koch{\textquotesingle}s curve. The Koch curve shares with the Brownian noise the properties of being continuous and nowhere differentiable, but the Koch curve is not a random fractal. Also the Koch curve is self\nobreakdash-similar while the Brownian process is self\nobreakdash-affine. Not all the curves used as examples in this paper are self\nobreakdash-similar or self\nobreakdash-affine, this is certainly the case of the uniform white noise, and probably also true for the rainfall curves in figure 1 or the mortality curve in figure 5. An interesting finding is also that equation (6) applied to sets of \textit{N{\textquotesingle}} samples may distinguish between waveforms that seemingly converge to the same value in the limit when ${N' \rightarrow \infty}$. This is the case of the uniform and Gaussian white noises. If \textit{x} and \textit{y} are \textit{U(0,1)} random variables, then ${\underset{N' \rightarrow \infty }\lim  f(x - y) \approx{N \left( \textit{0, 1/6} \right) }}$ and the expectation of ${\vert x-y \vert}$ is ${\approx \textit{1/3}}$. Representing the expectation by ${\mu}$, \textit{D} is then:
		
		\begin{equation}
			{\selectlanguage{english}
			D=\underset{N' \rightarrow \infty }{\lim } \left \{ 1+ \frac{ \ln \left[N'\cdot \sqrt{\mu ^{2}+\frac{1}{{N'}^2}} \right] }{ \ln (2 \cdot N')} \right \} = 2}
		\end{equation}
		
		With ${N' < \infty}$ equation (15) accurately predicts the values obtained experimentally in this communication. Yet the values of \textit{D} obtained for the Gaussian white noise with ${N' < \infty}$ differ significantly from those of the uniform white noise and from the prediction of equation (15) letting ${\mu =2/\sqrt{\pi }}$ the expectation of ${\vert x-y \vert}$ if \textit{x} and  \textit{y} are \textit{N(0,1)} random variables. The reason for this discrepancy is due to the fact that the length a curve made by adding \textit{N} segments of length ${\vert x-y \vert}$ is ${N \cdot \mu}$  only if \textit{x} and \textit{y} are bound in some finite interval \textit{(a,b)}. If \textit{x} and \textit{y} are \textit{N(0,1)} random variables, their difference is a \textit{N(0,2)} random variable with finite probability of getting close to infinitely large in absolute value at any time and thus we may expect that:

		\begin{equation*}
			\overset{N'}{\underset{i=1}{\sum }}{|x_{{i}}-y_{{i}}|}\neq N\cdot \mu
		\end{equation*}

		for any ${\textit{N'} < \infty}$. When ${\vert x-y \vert}$ gets very large the remaining points (and their distances) are rescaled to smaller values in the embedding unit square. This determines a smaller covering of the square by the curve and a smaller fractal dimension, as observed in table 2 for ${N' < \infty}$.
		
The value of \textit{D} calculated with equation (6), most likely underestimates to some degree ${D_H}$ of a waveform that is sampled with ${N' \rightarrow \infty}$. And yet, the values presented in table 2 indicate that the underestimation is < 5\%,  when you know the exact value of ${D_H}$. This is very good for many practical situations. The data also demonstrates that \textit{D} increases monotonically towards ${D_H}$ as ${N'}$ increases for all curves in table 2, and also that for all values ${1 < D < 2}$. For the case of the Koch's triadic curve ${D \rightarrow D_H}$ when ${N' \rightarrow \infty}$; this is only to be expected since \textit{D} is an approximate form of ${D_H}$ calculated for waveforms embedded in a unit square. Thus the results, show that all curves in table 2 are fractal in Manderbrot's sense; in this, there is perfect qualitative agreement with all the methods that are based on Hurst's exponent determination.

	\section[5. Acknowledgements]{5. Acknowledgements}
			
		Its Administrator, Mr. David Jones, kindly provided the data on rainfall in the Turagua Ranch. Dr. Yajaira Freites (Department on Science Studies, IVIC) brought \ to my attention the data on the Spanish Flu epidemics in 1918 and provided demographic information of the Federal District in 1920. Some software and hardware used were financed by a grant from Siemens\nobreakdash-Elema, A.B. Solna, Sweden. The author is indebted to an unknown reviewer whose comments led to equation (15) and its discussion.

		\bigskip
		
		\section[6. Appendix ]{6. Appendix }
		
		\subsection[6.1 A simple algorithm to calculate the fractal dimension of a waveform]{6.1 A simple algorithm to calculate the fractal dimension of a waveform}
		
		\bigskip
		
		{\selectlanguage{english}
		{ A simple algorithm in
		QuickBASIC}{\textsuperscript{TM}}{
		(Microsoft Corp., Seattle, WA) to calculate
		}{\textit{D}}{ is
		listed below. The choice of programming language is due to the ready
		availability of the compiler in recent operating systems from Microsoft
		and IBM. The algorithm is structured to ease translation to formal
		languages such as Pascal, C or C++.}}

		\bigskip
		
		{\selectlanguage{english}
		{\textquotesingle} ************ Program to Calculate Fractal Dimension
		of Waveforms }
		
		{\selectlanguage{english}
		\ \ \ \ \ DECLARE SUB LenCalc (y!(), ymin!, ymax!, N\%, Length!) }
		
		{\selectlanguage{english}
		\ \ \ \ \ DECLARE SUB DataInput (x!, y!, N\%) }
		
		{\selectlanguage{english}
		\ \ \ \ \ DIM x!(300), y!(300) }
		
		{\selectlanguage{english}
		\ \ \ \ \ CLS }
		
		{\selectlanguage{english}
		\ \ \ \ \ PRINT {\textquotedbl}Fractal Dimension of
		Waveforms{\textquotedbl}: PRINT }

		\bigskip
		
		{\selectlanguage{english}
		\ \ \ \ \ PRINT {\textquotedbl}Steps (N){\textquotedbl}, {\textquotedbl}
		\ x{\textquotedbl}, {\textquotedbl} \ y{\textquotedbl}, {\textquotedbl}
		\ D{\textquotedbl}: PRINT }
		
		{\selectlanguage{english}
		\ \ \ \ \ \ \ \ {\textquotesingle} ******** Get Initial Values
		************** }
		
		{\selectlanguage{english}
		\ \ \ \ \ N\% = 1 }
		
		{\selectlanguage{english}
		\ \ \ \ \ Length! = 0 }
		
		{\selectlanguage{english}
		\ \ \ \ \ CALL DataInput(x!(N\%), y!(N\%), N\%) }
		
		{\selectlanguage{english}
		\ \ \ \ \ PRINT, {\textquotedbl} \ {}-, }
		
		{\selectlanguage{english}
		\ \ \ \ \ ymax! = y!(1) }
		
		{\selectlanguage{english}
		\ \ \ \ \ ymin! = y!(1) }
		
		{\selectlanguage{english}
		\ \ \ \ \ \ \ \ {\textquotesingle} *** Loop to Calculate Fractal
		Dimension **** }
		
		{\selectlanguage{english}
		\ \ \ \ \ DO }
		
		{\selectlanguage{english}
		\ \ \ \ \ \ \ \ N\% = N\% + 1 }
		
		{\selectlanguage{english}
		\ \ \ \ \ \ \ \ CALL DataInput(x!(N\%), y!(N\%), N\%) {\textquotesingle}
		***** Data enter here ***** }
		
		{\selectlanguage{english}
		\ \ \ \ \ \ \ \ IF (y!(N\%) {\textgreater}= ymax!) THEN ymax! = y!(N\%)
		}
		
		{\selectlanguage{english}
		\ \ \ \ \ \ \ \ IF (y!(N\%) {\textless}= ymin!) THEN ymin! = y!(N\%) }
		
		{\selectlanguage{english}
		\ \ \ \ \ \ \ \ CALL LenCalc(y!(), ymin!, ymax!, N\%, Length!) }
		
		{\selectlanguage{english}
		\ \ \ \ \ \ \ \ D! = 1 + (LOG(Length!)-LOG(2)) / LOG(2*(N\% {}- 1)) }
		
		{\selectlanguage{english}
		\ \ \ \ \ \ \ \ PRINT, D! }
		
		{\selectlanguage{english}
		\ \ \ \ \ LOOP WHILE (N\% {\textless}= 300) }
		
		{\selectlanguage{english}
		END \ \ \ \ {\textquotesingle} ***** End of Main Program ***** }

		\bigskip
		
		{\selectlanguage{english}
		SUB DataInput (x!, y!, N\%) \ {\textquotesingle} ***** Subroutine for
		Data Input ***** }
		
		{\selectlanguage{english}
		\ \ \ \ PRINT N\%; }
		
		{\selectlanguage{english}
		\ \ \ \ PRINT , ; }
		
		{\selectlanguage{english}
		\ \ \ \ INPUT ; x! }
		
		{\selectlanguage{english}
		\ \ \ \ PRINT , ; }
		
		{\selectlanguage{english}
		\ \ \ \ INPUT ; y! }
		
		{\selectlanguage{english}
		END SUB \ \ \ {\textquotesingle} ***** End of Data Input ***** }

		\bigskip
		
		{\selectlanguage{english}
		\ \ {\textquotesingle} ****** \ \ \ LenCalc; Subroutine that Calculates
		the Normalized Length of the Waveform }
		
		{\selectlanguage{english}
		SUB LenCalc (y!(), ymin!, ymax!, N\%, Length!) }
		
		{\selectlanguage{english}
		\ \ \ \ IF N\% = 1 THEN }
		
		{\selectlanguage{english}
		\ \ \ \ \ \ \ PRINT, {\textquotedbl} \ {}-{\textquotedbl} }
		
		{\selectlanguage{english}
		\ \ \ \ ELSE }
		
		{\selectlanguage{english}
		\ \ \ \ \ \ \ Length! = 0 }
		
		{\selectlanguage{english}
		\ \ \ \ \ \ \ FOR i\% = 1 TO N\% }
		
		{\selectlanguage{english}
		\ \ \ \ \ \ \ \ \ \ \ y! = (y!(i\%) {}- ymin!) / (ymax! {}- ymin!) }
		
		{\selectlanguage{english}
		\ \ \ \ \ \ \ \ \ \ \ IF (i\% {\textgreater} 1) THEN Length! = Length! +
		SQR((y! {}- yant!) \^{} 2 + (1! / (N\% {}- 1)) \^{} 2) }
		
		{\selectlanguage{english}
		\ \ \ \ \ \ \ \ \ \ \ yant! = y! }
		
		{\selectlanguage{english}
		\ \ \ \ \ \ \ NEXT i\% }
		
		{\selectlanguage{english}
		\ \ \ \ END IF }
		
		{\selectlanguage{english}
		END SUB \ \ {\textquotesingle} ***** End of LenCalc ***** }

		\bigskip

		\bigskip
	
	\section[7  References]{7  References}
	
	\bigskip
	
	{\selectlanguage{english}
	{\textbf{1}}{  Badii, R. and Politi, A. (1985) Statistical description of chaotic attractors: The dimension function.}{\textit{ J. Stat. Phys.}}{ 40: 725\nobreakdash-750.}}

	\bigskip
	
	{\selectlanguage{english}
	{\textbf{2}}{ Barnsley, M.F. (1993) }{\textit{Fractals Everywhere}}{. Ch. II and III, Academic Press Professional, Cambridge MA.}}

	\bigskip
	
	{\selectlanguage{english}
	{\textbf{3}}{ Box, G.E.P. \ and Muller, M.E. (1958) A Note on the Generation of Random Normal Deviates. }{\textit{Ann.Math.Statist.}}{ 22: 610\nobreakdash-611.}}

	\bigskip
	
	{\selectlanguage{spanish}
	{4 Brito\nobreakdash-Figueroa, F. (1974) }{\textit{Historia Econ\'omica y Social de Venezuela}}{\textbf{Vol. II}}{ Ediciones de La Biblioteca, Universidad Central de Venezuela, Caracas.}}

	\bigskip
	
	{\selectlanguage{english}
	{\textbf{5}}{ Colquhoun, D. (1971) }{\textit{Lectures on biostatistics}}{. p 49. Claredon Press, Oxford.}}

	\bigskip
	
	{\selectlanguage{english}
	{\textbf{6}}{ Dahlquist, G. and \ Bj\"ork, {\AA}. (1974) }{\textit{Numerical Methods}}{. Prentice Hall, Inc., Englewood Cliffs.}}

	\bigskip
	
	{\selectlanguage{spanish}
	{\textbf{7}}{ D\'avila, D. (1995) }{\textit{Con el Pa\~nuelo en la Nariz (Caracas y la Influenza de 1918).}}{ }{\textit{Historia para Todos. 8.}}{ Historiadores SC, Caracas.}}

	\bigskip
	
	{\selectlanguage{english}
	{\textbf{8}}{ Elbert, T., Ray, W.J. ,Kowalik, Z.J., Skinner, J.E. , Graf, K.E. \ and Birbaumer, N. (1994) Chaos and Physiology: Deterministic Chaos in Excitable Cell Assemblies. }{\textit{Phys.Rev}}{. 74: 1\nobreakdash- 47.}}

	\bigskip
	
	{\selectlanguage{english}
	{\textbf{9}}{ Grassberger, P. and Procaccia, I. (1983a) Estimation of the Kolmogorov entropy from a chaotic signal.
	}{\textit{Phys.Rev. A}}{, 28: 2591\nobreakdash-2593.}}

	\bigskip
	
	{\selectlanguage{english}
	{\textbf{10}}{ Grassberger, P. and Procaccia, I. (1983b) Measuring strangeness of strange attractors. }{\textit{Phys. Lett.
	D}}{ 148: 63\nobreakdash-68.}}

	\bigskip
	
	{\selectlanguage{english}
	{\textbf{11}}{ Grassberger, P. and Procaccia, I. (1983c) Characterization of strange attractors }{\textit{Phys. Rev. Lett.}}{ 50: 346\nobreakdash-349.}}

	\bigskip
	
	{\selectlanguage{english}
	{\textbf{12}}{ Hastings, H.H. and Sugihara, G. (1993) }{\textit{Fractals. A User{\textquotesingle}s Guide for the Natural Sciences}}{. Oxford University Press, Oxford.}}

	\bigskip
	
	{\selectlanguage{english}
	{\textbf{13}}{ Hurst, H.E. (1951) Long\nobreakdash-term storage capacity of reservoirs. }{\textit{Trans.Am.Soc. Civil Eng.}}{
	116}{\textbf{:}}{ 770\nobreakdash-808.}}

	\bigskip
	
	{\selectlanguage{english}
	{\textbf{14}}{ Hurst, H.E. (1956) Methods of using long\nobreakdash-term storage in reservoirs. }{\textit{Proc.Inst. Civil
	Eng.}}{\textbf{ 5, }}{Part 1}{\textbf{:}}{ 519\nobreakdash-577.}}

	\bigskip
	
	{\selectlanguage{english}
	{\textbf{15}}{ Katz, M.J. (1988) Fractals and the analysis of waveforms. }{\textit{Comput.Biol.Med.}}{ 18: 145.}}

	\bigskip
	
	{\selectlanguage{english}
	{\textbf{16}}{ Kirkpatrick, S. and Stoll, E.P. \ (1981) A very fast shift\nobreakdash-register sequence random number generator. }{\textit{J.Comp.Phys.}}{ 40: 517\nobreakdash-526.}}

	\bigskip
	
	{\selectlanguage{english}
	{\textbf{17}}{ Mandelbrot, B.B. (1983) }{\textit{The Fractal Geometry of Nature.}}{ \ W.H. Freeman and Co., New York.}}

	\bigskip
	
	{\selectlanguage{english}
	{\textbf{18}}{ Mandelbrot, B.B. \ (1986) Fractals and the rebirth of iteration theory. In: }{\textit{The beauty of fractals}}{ by H.\nobreakdash-O. Peitigen and P.H. Richter, p 150. Springer Verlag, Berlin.}}

	\bigskip
	
	{\selectlanguage{english}
	{\textbf{19}}{ Nicolis, G. and Prigoyine, I. \ (1989) }{\textit{Exploring Complexity. An Introduction.}}{ W.H. Freeman and Co., New York.}}

	\bigskip
	
	{\selectlanguage{english}
	{\textbf{20}}{ Pincus, S.M., Gladstone, I.M. \ and Ehrenkranz, R.A. \ (1991) A regularity statistics for medical data analysis. }{\textit{J.Clin.Monit.}}{ }{\textbf{7}}{: 335\nobreakdash-345.}}

	\bigskip
	
	{\selectlanguage{english}
	{\textbf{21}}{ Pocaterra, J.R. (1990) }{\textit{Memorias de un Venezolano de la Decadencia}}{. Vol. 1, pg 237. Biblioteca Ayacucho, Caracas.}}

	\bigskip
	
	{\selectlanguage{english}
	{\textbf{22}}{ Press, W.H., Teukolsky, S.A., Vetterling, W.T. and Flannery, B.P. (1992) }{\textit{Numerical Recipes in C. The art of Scientific Computing}}{. Cambridge University Press: New York.}}

	\bigskip
	
	{\selectlanguage{english}
	{\textbf{23}}{ Schroeder, M. (1991) }{\textit{Fractals, Chaos and Power Laws.}}{ Freeman and Co., New York.}}

	\bigskip
	
	{\selectlanguage{english}
	{\textbf{24}}{ Tsonis, A. (1992) }{\textit{Chaos: From Theory to Applications.}}{ Plenum, New York.}}

	\bigskip
	
	{\selectlanguage{english}
	{\textbf{25}}{ Wilks, S.S. (1962) }{\textit{Mathematical Statistics.}}{ John Wiley \& Sons, New York.}}
\end{document}